Anisotropic electronic transport of the two-dimensional electron system in $Al_2O_3$/$SrTiO_3$ heterostructures

K. Wolff,[1] R. Schäfer,[1] M. Meffert,[2] D. Gerthsen,[2] R. Schneider,[1] and D. Fuchs[1]

[1]Karlsruher Institut für Technologie, Institut für Festkörperphysik, 76021 Karlsruhe, Germany
[2]Karlsruher Institut für Technologie, Laboratorium für Elektronenmikroskopie, 76021 Karlsruhe, Germany

Transport measurements on the two-dimensional electron system in $Al_2O_3$/$SrTiO_3$ heterostructures indicate significant non-crystalline anisotropic behavior below $T \approx 30$ K. Lattice dislocations in $SrTiO_3$ and interfacial steps are suggested to be the main sources for electronic anisotropy. Anisotropic defect scattering likewise alters magnetoresistance at low temperature remarkably and influences spin-orbit coupling significantly by the Elliot-Yafet mechanism of spin relaxation resulting in anisotropic weak localization. Applying a magnetic field parallel to the interface results in an additional field–induced anisotropy of the conductance, which can be attributed to Rashba spin-orbit interaction. Compared to $LaAlO_3$/$SrTiO_3$, Rashba coupling seems to be reduced indicating a weaker polarity in $Al_2O_3$/$SrTiO_3$ heterostructures.

I. INTRODUCTION

Two-dimensional electron systems (2DES) at the interface between insulating oxides have gained huge interest in the last years. The importance for multifunctional all-oxide devices as well as the intriguing properties of strongly correlated and confined 2DES gave rise to many interesting scientific works. The emergence of superconductivity [1], multiple quantum criticality [2], tunable spin-orbit coupling (SOC) [3], and magnetism [4] at the interface between $LaAlO_3$ and $TiO_2$-terminated $SrTiO_3$ (LAO/STO) have made STO-based heterostructures a prototypical system for studying low-dimensional strongly correlated electron systems. Charge carriers in 2DES of STO based heterostructures display Ti $3d$-derived $t_{2g}$-orbital character extending over a few STO layers from the interface [5]. The broken inversion symmetry at the interface results in a splitting of the $t_{2g}$ bands into a lower $d_{xy}$ singlet and an upper $d_{xz}$, $d_{yz}$ doublet state, where the $z$-direction is along the surface normal. The band filling strongly depends on sheet carrier density $n_s$, suggesting a Lifshitz transition at $n_c \approx 1.7 \times 10^{13}$ cm$^{-2}$ [6]. For $n_s > n_c$ most of the charge carriers accumulate in the $d_{xz}$, $d_{yz}$ bands [5].

The polar discontinuity at the LAO/STO interface leads to a rather strong Rashba-type SOC [7] and is considered to play an important role with respect to interfacial conductivity [8]. However, oxygen vacancies may also act as a possible source for charge carrier doping of STO [9]. For example, chemical redox reactions at the interface between STO and other complex oxides provide an alternative approach to create new types of 2DES in complex oxide heterostructures [10,11], where 2D metallic behavior in e. g., amorphous aluminum oxide/STO heterostructures, is assumed to be dominated by oxygen vacancies and is not provided by electronic reconstruction.

The metallic interface between strongly disordered and quasi amorphous aluminum oxide, grown at low substrate temperature, $T_s \leq 200°C$, and (001) oriented, TiO$_2$-terminated STO (AO/STO) displays sheet carrier density, Hall mobility, and even superconducting properties which are well comparable to those of epitaxially grown LAO/STO [12]. Further motivation for using AO/STO is the low deposition temperature which is very advantageous with respect to technical, large scale production and processing. Close to the superconducting transition, Van-der-Pauw resistance measurements on AO/STO indicate anisotropic electronic transport. Anisotropic striped, filamentary electronic structure due to mesoscopic inhomogeneities has been observed alike in the 2DES of epitaxial LaTiO$_3$/STO [2] and LAO/STO [13-17]. On the one side, extrinsic defects and impurities, or a net surface charge at step edges [18] appear to be mainly responsible for the electric inhomogeneity. On the other side, strong Rashba coupling may also lead to charge segregation and intrinsic electronic phase separation even in perfectly clean and homogeneous LAO/STO [19]. Therefore, more detailed transport measurements with respect to anisotropic electronic behavior are necessary for a better understanding of emerging nonlocal resistance phenomena in 2DES of STO-based heterostructures.

In this paper, we report on transport measurements on AO/STO microbridges patterned along different in-plane crystallographic directions using STO substrates with different step edge alignments. Lattice dislocations in STO and interfacial steps appear as the main sources for electronic anisotropy, likewise influencing SOC and magnetoresistance. An in-plane magnetic field results in Rashba-induced oscillations of the conductance. The Rashba coupling seems to be reduced compared to LAO/STO indicating weaker polarity in AO/STO.

## II. EXPERIMENTAL

In order to characterize anisotropic electronic transport of the 2DES in AO/STO, microbridges with length of 100 µm and width of 20 µm in Hall bar geometry were patterned along specific crystallographic directions using a $CeO_2$ hard mask technique [20], see Fig. 1 (a). The microbridges are labeled alphabetically from A to E, and display angle $\varphi$ = 0°, 22.5°, 45°, 67.5°, and 90°, towards the [100] direction, respectively. The deposition of $CeO_2$ as well as the subsequent ablation of $Al_2O_3$ in order to form the 2DES at the interface of AO/STO was carried out by pulsed laser deposition on (001) oriented $TiO_2$-terminated STO substrates. Contacts to the buried 2DES were produced by ultrasonic Al-wire bonding. In the used current- and temperature-regime the contacts showed clear Ohmic behavior. More details with respect to sample preparation are given in references [12,20]. The single-type termination of the STO substrates usually leads to the formation of a stepped surface with a step-height of one STO unit cell [21]. Motivated by previous observations of the possible influence of interfacial steps on the anisotropic transport behavior [22,12] we used substrates with different step edge alignment with respect to the [100] direction. For sample I, the angle between the step edges and the [100] direction amounts to $\omega \approx 85°$ and for sample II $\omega \approx 55°$. The surface topography of sample I and II is shown in Fig. 1 (b). All the used substrates came from the same batch (CrysTec company), hence, displaying the same crystalline quality. The miscut angle of the substrates typically amounts to 0.1° - 0.2° which results in a terrace-width of 100 – 250 nm (see Fig. 1 (b)).

Measurements of the sheet resistance $R_s$ were carried out in a physical property measurement system (PPMS) from Quantum Design in the temperature and magnetic field ranges $2 \text{ K} \leq T \leq 300 \text{ K}$ and $0 \leq B \leq 14 \text{ T}$. In order to avoid charge carrier activation by light [23,24], alternating current measurements ($I_{ac}$ = 3 µA) were started not before 12 hours after loading the samples to the PPMS. The magnetoresistance, MR = $[R_s(B) - R_s(0)]/R_s(0)$, was measured with magnetic field normal and parallel to the interface. For measuring $R_s(B)$ with rotating in-plane magnetic field $B_{ip}(\phi)$, a sample rotator was used. The angle $\phi$ between $B_{ip}$ and [100]-direction was varied from 0° - 360°. Special care was taken to minimize sample wobbling in the apparatus. Residual tilts (1° - 2°) of the surface normal with respect to the rotation axis which produces a perpendicular field component oscillating in sync with $\phi$ could be identified by comparison of $R_s(B,\phi)$ for different microbridges and was therefore corrected properly.

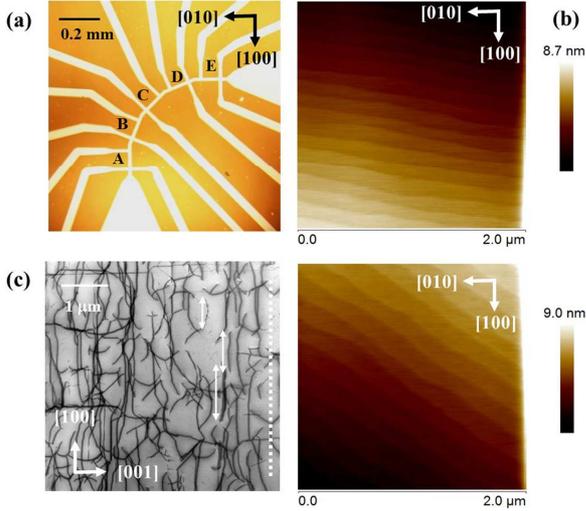

FIG. 1. (a) Micrograph of a patterned sample. Sharp contrast between AO/CeO$_2$ (dark) and AO/STO (bright) enables identifying microbridges labeled alphabetically from A – E. (b) Surface topography before Al$_2$O$_3$ deposition characterized by atomic force microscopy shown for sample I ($\omega \approx 85°$) (top) and for sample II ($\omega \approx 55°$) (bottom). The images were taken on microbridge A. Step-edge orientation was found to be the same for all microbridges. (c) Cross-sectional bright-field scanning transmission electron microscopy image of a standard STO substrate. The dark lines are dislocations which are present with a high density on (001) lattice planes, i.e., parallel to the interface plane indicated by the dotted line. The white arrows mark dislocation segments with finite length.

## III. RESULTS AND DISCUSSION

### A. Temperature dependence of the electronic transport

First, we report on sheet resistance measurements as a function of temperature without application of a magnetic field. Fig. 2 displays the sheet resistance $R_s$ versus $T$ of the microbridges A – E of sample I ($\omega \approx 85°$) and sample II ($\omega \approx 55°$). For 100 K $\leq T \leq$ 300 K, both samples display nearly identical $R_s(T)$. $R_s(T)$ shows isotropic behavior with an approximate $T^2$-dependence. Such a $T$-dependence is often observed in STO-based heterostructures and $n$-type doped bulk STO [25,26] and attributed to electron-phonon scattering. Cooling-down results in a shallow minimum around 30 K below which $R_s$ increases. For $T < 10$ K $R_s$ is nearly constant indicating dominant $T$-independent impurity scattering. The resistivity ratio between 300 K and 10 K amounts to about 20. However, for $T$

< 5 K $R_s$ slightly decreases again for some bridges. It is very likely, that this behavior is caused by weak antilocalization [27,28], a well- known feature of 2DES displaying SOC such as LAO/STO [3]. We will discuss this point in more detail later. Obviously, electronic transport becomes anisotropic below 30 K, where impurity scattering starts to dominate $R_s(T)$ [26]. For sample I $R_s(T = 5\ \text{K})$ is lowest for bridge E where the interfacial step edges are aligned nearly parallel to the macroscopic current direction. In comparison, current path oriented nearly perpendicular to interfacial step edges (bridge A) results in an increase of $R_s(5\ \text{K})$ by about 17%. Bridge C, where current path is close to the [110] direction, displays the highest $R_s$, about 30 % higher compared to $R_s$ of bridge E. For sample II, the anisotropic behavior for $T < 30$ K is less pronounced. Although the angles between microbridges and step edges are quite different for sample I and sample II, $R_s$ is again maximal for bridge C. However, the variation of $R_s$ from bridge A to E is only about 4%.

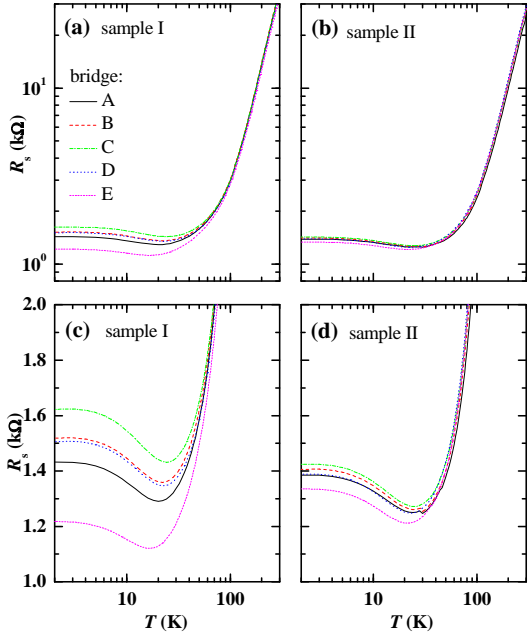

FIG. 2. Sheet resistance $R_s$ of (a) sample I and (b) sample II versus $T$ as obtained from 4 point measurements on the microbridges A – E. The distinct anisotropic behavior of $R_s$ at low $T$ is visualized in the semi-logarithmic plots (c) and (d). For sample I interfacial steps are running nearly parallel to bridge E ($\omega \approx 85°$), whereas for sample II terraces are running close to the [110] direction ($\omega \approx 55°$).

The ratio $R_s$(bridge C)/$R_s$(bridge E) (≈ 1.33 for sample I) changes only little from 10 K to 2 K, indicating that the dominant contribution to the anisotropy of $R_s$ is caused by impurity or defect scattering. Additional contributions by, e. g., quantum effects such as weak localization (WL) or electron-electron interaction (EEI) cannot be excluded and may be present as well but are suggested to be less important than impurity scattering.

On the one side, impurity or defect scattering of 2DES in STO-based heterostructures is caused by the same mechanism as in the STO bulk. Fig. 1 (c) shows a cross-section image of a standard STO substrate taken by bright-field scanning transmission electron microscopy. The image displays a high density of dislocations on the (001) lattice plane, i.e., parallel to the interface indicated by the dotted line. The projected length of the dislocation lines varies significantly along the [010] viewing direction (some short dislocation segments are marked by white arrows in Fig. 1 (c)), indicating that the dislocation lines are oriented along different directions on the (001) lattice planes.

Flame fusion (Verneuil) -grown STO single crystals are indeed well known for displaying high dislocation densities (> $10^6$ cm$^{-2}$) [29]. Most prominent are <110> lattice dislocations with preferential {1-10} slip planes leading, for instance, to an atypical mechanical (plastic) behavior [30]. Such <110> dislocations also cause charge carrier scattering and therefore increased resistance for the perpendicular current direction. An anisotropic distribution of dislocation lines along [110] and [1-10] direction, is expected to result in an anisotropy of $R_s$.

On the other side, defect scattering at the interface has to be taken into account as well. Interfacial steps likely decrease charge carrier mobility and may increase low-temperature resistance in LAO/STO heterostructures [22]. For both samples I and II we find that $R_s$ is higher when the current is perpendicular to the step edges and lower when the current is parallel. Interfacial steps may also result in further break up of inversion symmetry within the film plane resulting in SOC, in addition to what usually results from symmetry breaking perpendicular to the interface [31].

In the following, we have modeled the in-plane anisotropy of $R_s$ at $T = 5$ K for sample I and II by considering anisotropic contributions to the resistance originating from charge carrier scattering by inhomogeneous distribution of dislocation lines along [110] and [1-10] direction ($r_d$) and interfacial steps and terraces ($r_t$) resulting in a total sheet resistance $R_s(\varphi) = r_0 + r_d(\varphi) + r_t(\varphi)$ with $r_d(\varphi) = \hat{r}_d \times \sin(\varphi-\omega_d)^2$ and $r_t(\varphi) = \hat{r}_t \times \sin(\varphi-\omega_t)^2$. $r_0$ represents isotropic contributions to $R_s$ from, e. g., point defects, $\hat{r}_d$ and $\hat{r}_t$ the amplitudes of $r_d(\varphi)$ and $r_t(\varphi)$, $\omega_d$ the angle between the preferential direction of dislocation lines and the [100] direction, and $\omega_t = 85°$ (sample I) or $\omega_t = 55°$ (sample II). The angular dependence of $R_s(\varphi)$ as well as the isotropic part $r_0$ and the anisotropic parts $r_d$ and $r_t$ are shown for sample I and II in Fig. 3. The model described above results in a consistent description of $R_s(\varphi)$ for samples displaying different step edge alignment. The isotropic part $r_0$ of sample II is somewhat larger compared to sample I which, however, may be related to a more homogeneous distribution of dislocations in the sample, documented by a smaller $\hat{r}_d$ in comparison to sample I. The maximum anisotropy, $(R_s^{max}-R_s^{min})/R_s^{min}$, as deduced from the maximum ($R_s^{max}$) and minimum values ($R_s^{min}$) of $R_s(\varphi)$ amounts to 55% and 18.5% for sample I and II, respectively. The decrease of the anisotropy in sample II seems to result mainly from a more isotropic distribution of dislocation lines and hence smaller $\hat{r}_d$. For better comparison, the anisotropic contributions $r_d(\varphi)$ and $r_t(\varphi)$ of both samples are plotted in Fig. 3 (b) and (d).

To proof anisotropic behavior of $r_d(\varphi)$ again, we likewise prepared microbridges with $\varphi = 45°$ -135°. The samples all displayed clear anisotropy of $R_s$(5 K) with respect to the [110] and [1-10] direction, i. e., $R_s(\varphi =45°) > R_s(90°) > R_s(135°)$ or $R_s(45°) < R_s(90°) < R_s(135°)$.

Because of the different alignment of the step edges, the minima of $r_t(\varphi)$ are shifted from $\varphi = 85°(265)°$ for sample I to 55°(235°) for sample II. Obviously, $\hat{r}_t$ of sample II is reduced compared to sample I. The larger terrace width of sample II (*cf.* Fig. 1 (b)) results in a lower step density and therefore in a reduced $\hat{r}_t$. The ratio of $\hat{r}_t$ between sample I and II ($\approx$ 1.6) compares nearly perfectly with the inverse ratio of the terrace width of both samples, which strongly supports our model.

The mean free electron path, $\lambda_{mfp}$, can be deduced from the two-dimensional Fermi velocity $v_F = \hbar(2\pi n_s)^{1/2}/m^*$ and the relaxation time $\tau = \mu \times m^*/e$ by : $\lambda_{mfp} = v_F \times \tau$, where $\hbar$ is the Planck constant divided by $2\pi$, $m^*$ the effective electron mass, $\mu$ the Hall mobility and $e$ the elementary charge. With $n_s = 2\times10^{13}$ /cm$^2$ and $\mu = 200$ cm$^2$/(Vs) at $T = 5$ K (see also next section) this results in $\lambda_{mfp} \approx 15$ nm. Compared to the terrace width $w$ (100 – 250 nm) $\lambda_{mfp}$ is quite small, only 8 - 15 %. Therefore, the influence of the step edges upon scattering rate and total $R_s$ is expected to be rather small, too. In comparison to the total $R_s$, $\hat{r}_t$ indeed only amounts to about 5.6% for sample I and 8.6% for sample II. However, one has to be aware, that $\hat{r}_t$ only accounts for surface scattering effects in contrast to $r_0$ and $\hat{r}_d$ which comprise electron scattering perpendicular to the interface as well. Hence, direct extraction or comparison of scattering rates from $r_0$, $\hat{r}_d$, and $\hat{r}_t$ might be critical. Obviously, dominant contributions to the anisotropic behavior of $R_s(\varphi)$ at low $T$ originate from defect scattering by bulk-like dislocations, being about 2 - 4 times larger in amplitude compared to interfacial scattering by step edges. The increased factor may be inherently related to the ratio between the thickness of the 2DES [32] and the interfacial steps. The largest contribution to $R_s$ is the isotropic part $r_0$.

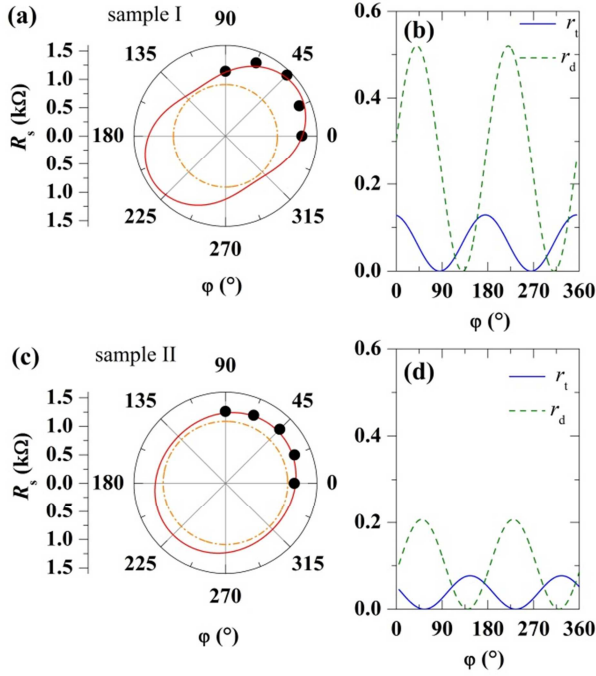

FIG. 3. Polar plots of the angular dependence of $R_s(\varphi)$ at $T = 5$ K (symbols) for (a) sample I where step edges are aligned nearly perpendicular to the [100] direction ($\omega_t \approx 85°$) and (c) for sample II where step edges are aligned by $\omega_t \approx 55°$ with respect to the [100] direction. $\varphi$ is the in-plane angle between current- and [100] direction. The total sheet resistance is modeled by $R_s(\varphi) = r_0 + r_d(\varphi) + r_t(\varphi)$ (solid line), see text. The isotropic part $r_0$ is shown by dashed dotted line. For better comparison, anisotropic contributions $r_d(\varphi)$ (dashed line), caused by inhomogeneous distribution of <110> dislocation lines ($\omega_d = 135°$), and $r_t(\varphi)$ (solid line) caused by the terraces are displayed for sample I and II in (b) and (d), respectively.

## B. Magnetic field dependence of the electronic transport

Measurements of the electronic transport were carried out with the magnetic field $B$ normal and parallel to the conducting interface. We first report on the Hall and MR measurements where $B$ was applied normal to the 2DES. The Hall resistance ($R_{xy}$) was measured in the temperature and magnetic field ranges 2 K $\leq T \leq$ 300 K and $0 \leq B \leq 14$ T. $R_{xy}(B)$ shows isotropic behavior with respect to φ for $T > 100$ K and a linear field dependence, suggesting dominating single-type (electron-like) carrier transport. Multiple-type carrier transport, i. e., a nonlinear field dependence of $R_{xy}(B)$ appears below about 30 K where also some anisotropic behavior becomes evident. A two-band model is often used in the LAO/STO literature to extract the mobilities and densities when Hall resistance traces are S-shaped. However, this model assumes that the parameters of the bands are independent of $B$. Therefore, it should be used with caution in case of LAO/STO [6], where it might lead to large mistakes in some of the extracted parameters. In the following, we used only the robust predictions of this model, i. e., the asymptotic value of $R_{xy}$ at high fields and in the limit of zero giving the total sheet carrier density $n_{tot}$ and the sheet carrier density of the charge carriers having the highest mobility, $n_{hi}$, respectively. Note, that $n_{hi}$ differs only by about 10% from $n_{tot}$. Hence, charge carriers with lower mobility obviously contribute only less to the electronic transport. For that reason we concentrate on discussing only the impact of the charge carriers with the highest mobility $n_{hi}$ on the electronic transport. $n_{tot}$ is displayed versus $T$ in Fig. 4 (a). Data are shown for bridge A to E of sample I. Data for sample II (not shown) are very similar. At 300 K $n_{tot}$ amounts to about $4 \times 10^{13}$ cm$^{-2}$ and drops down to $\approx 2.5 \times 10^{13}$ cm$^{-2}$ for $T \leq 10$ K. The $T$-dependence of $n_{tot}$ is typical for 2DES in STO based heterostructures [26] and is usually interpreted as a freeze-out of charge carriers [33,34]. The Hall mobility of $n_{hi}$ was calculated by $\mu = (R_s(B = 0) \times n_{hi} \times e)^{-1}$, where $e$ is the elementary charge. The $T$-dependence of μ for bridge A to E of sample I is shown in Fig. 4 (b). In accordance with $R_s$, μ increases nearly proportional to $T^{-2}$ with decreasing $T$ due to the decrease of electron-phonon scattering. The highest mobility is obtained for $T \approx 20$ K amounting to about 350 cm$^2$/Vs. For $T \leq 10$ K, μ is limited by defect or impurity scattering as indicated by the $T$-independent behavior. Interestingly, for $T \leq 10$ K μ displays significant anisotropy. μ is about twice as large for bridge E than for bridge C, which indicates much higher defect or impurity scattering along bridge C. Furthermore, the mobility of bridge E, with mean current path parallel to the step edges is likewise larger compared to bridge A being perpendicular to the step edges. These results are in good agreement with our modeling of $R_s(T = 5$ K, $B = 0)$ shown before revealing electron scattering by anisotropic distribution of defects as the primary source for the anisotropic mobility of the 2DES.

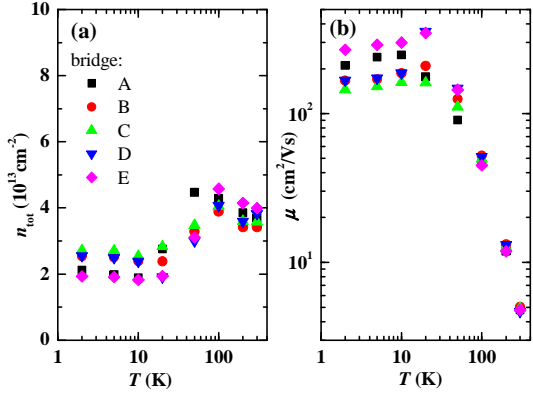

FIG. 4. (a) Total sheet carrier density $n_{\text{tot}}$ and (b) Hall mobility $\mu$ of the high mobility charge carriers versus $T$ for bridge A – E of sample I. $n_{\text{tot}}$ was deduced from the high field limit of $R_{xy}(B)$ and the mobility by $\mu = (R_s(B = 0) \times n_{\text{hi}} \times e)^{-1}$, see text.

The magnetoresistance MR of bridge A to E for sample I is shown for various temperatures in Fig. 5 (a). For $T \geq 100$ K the 2DES displays isotropic and small positive MR, less than 1%. However, cooling down results in a significant increase and a distinct anisotropic behavior of MR at $T = 10$ K.

A positive MR at low temperatures usually originates from the orbital motion of free carriers due to the Lorentz force, i. e., Lorentz (LZ) scattering. Assuming a two band model with different sheet carrier densities in both bands, the Hall mobility $\mu$ and cyclotron motion on open orbits, the field dependence of MR can be expressed by a Lorenzian function, i. e., the Kohler form: MR $\sim B^2/(1+(B/w)^2)$ [35]. The Lorenzian broadening $w$ strongly depends on the inverse of the mobility $\mu$. MR at 10 K can be perfectly described by classical LZ scattering mechanism, see fits (solid lines) to the data in Fig. 5 (a). The broadening $w$ which we deduce from the fits is nearly perfectly proportional to $\mu^{-1}$. The smallest broadening of MR is observed for bridge E displaying the highest mobility, whereas the largest broadening is obtained on bridge C showing the lowest $\mu$. Thus, the anisotropic behavior of MR is mainly caused by the variation of charge carrier mobilities (see Fig. 4 (b)).

Further cooling down to $T = 2$ K results in an additional contribution to the positive MR. However, significant changes to MR are restricted to $B < 8$ T, whereas for $B > 8$ T MR is well comparable to that at 10 K.

At low $T$ the magnetoresistance of a 2DES is usually dominated by contributions of electron-electron interaction and WL [28]. Previous studies show, that in LAO/STO heterostructures the breaking of inversion symmetry at the interface promotes Rashba-type spin-orbit interaction [3]. Therefore, in the diffusive regime of charge transport MR is well described by the 2D WL theory [27,36]. Zeeman corrections, which in case of LAO/STO are usually much smaller compared to the spin-orbit effects, were taken into account by Maekawa and Fukuyama (MF) [37]. Efforts were also made to extract the wave vector ($k$) dependence of spin splitting energy and Rashba effect from MR [38,39]. The $t_{2g}$ orbitals $d_{xz}$ and $d_{yz}$ derived from Ti 3$d$ states of STO lead to a $k^3$ spin splitting model (cubic Rashba effect), displaying similar field dependence of MR as deduced by MF [38]. In the following, the MF theory was used to fit the experimental data at 2 K in combination with a Kohler term as described above. The parameters of the MF-expression [3] are the inelastic field $B_i$, the spin-orbit field $B_{so}$, and the electron $g$-factor which enters into the Zeeman corrections.

The data at $T = 2$ K are perfectly described by the fits (see Fig. 5 (a)) allowing us to deduce $B_i$ and $B_{so}$. Zeeman corrections to MR were found to play only a minor role for $B \leq 14$ T. Separate contributions to MR from the Kohler term and the MF expression are shown in Fig. 5 (b) and (c), respectively. Obviously, contributions from classical LZ scattering mechanism at 2 K and 10 K are well comparable with respect to amplitude and broadening $w$. Anisotropic behavior of that part can therefore be explained in the same way as before, i. e., by a different charge carrier mobility of the micro-bridges.

Contributions from WL as deduced from the MF expression, are much "weaker" compared to contributions from Lorentz scattering at $B = 14$ T, i. e., |MR|≤ 2%, however dominate the total MR for $B < 2$ T. The maximum difference in conductance amounts to $\Delta\sigma = \sigma(B) - \sigma(0) \approx 0.3\ e^2/h$, verifying the WL correction. Here, $\sigma(B) = 1/R_s(B)$ and $h$ the Planck constant. Obviously, the various microbridges display different WL behavior alike. $B_i$ and $B_{SO}$ as deduced from the fits are shown in Fig. 6 (a) as a function of the in-plane direction of the current $0 \leq \varphi \leq 90°$. $B_i$ and $B_{SO}$ are in the range of 60 - 80 mT and 1 – 6 T, respectively. The values for $B_{so}$ for $\varphi \approx 0°$ and 90° are well comparable to those reported for LAO/STO samples with $\sigma(0) \approx 0.6$ mS [40] (sample I displays nearly the same mean conductance $\sigma(0) = 0.63$ mS). We therefore conclude that spin-orbit interaction in AO/STO is here controlled by Rashba effect alike. However, in contrast to $B_i$, which seems to depend very little on $\varphi$, $B_{SO}$ displays a distinct behavior on $\varphi$ and is mainly responsible for the anisotropic behavior of the WL contribution (see Fig. 5 (c)).

In case of Rashba coupling, the dephasing of electron spins, defined by the spin relaxation time $\tau_{SO} \propto 1/B_{SO}$, is described by D`yakonov-Perel (DP) mechanism of spin relaxation [41], leading to $\tau_{SO} \propto 1/\tau$, where $\tau$ is the elastic scattering time. For LAO/STO this seems to be fulfilled quite well [3]. Because of the symmetric band structure of (001) oriented STO-based heterostructures, Rashba coupling is expected to be isotropic and $B_{SO}$ should not depend on φ. However, interfacial steps may also result in a further break up of inversion symmetry within the film plane resulting in a change of Rashba type spin orbit coupling and hence $B_{SO}$ [31].

On the other side, the 2DES in STO-based heterostructures may also be sensitive to the Elliot-Yafet (EY) mechanism of spin relaxation [42-44]. The EY mechanism takes into account dephasing of spins by impurities, lattice defects, or phonons. In contrast to the DP mechanism, the EY mechanism leads to the Elliot relation, $\tau_{SO} \propto \tau$ [42,43].

In Fig. 6 (b) we have plotted the difference $\Delta B_{so} = B_{so}(\varphi) - B_{so}(\varphi=90°)$ as a function of the inverse of the Hall mobility (obtained for the different microbridges at φ = 0, 22.5, 45, 67.5 and 90°). $\Delta B_{so}$ obviously increases nearly linearly with $1/\mu$. In the diffusive regime of electron transport, the Drude model yields $\tau \propto \mu$ and hence, $\Delta B_{so} \propto 1/\tau$, indicating spin relaxation dominated by EY mechanism. So, in principle both types of spin relaxation are at work, DP and EY mechanism where EY mechanism probably contributes mainly to the spin-relaxation at φ ≈ 45°. The anisotropic distribution of defects as discussed in section A which lead to anisotropic behavior of μ(φ) evidently affects spin-orbit interaction and results in the emergence of anisotropic WL. The distinct relation $\Delta B_{so} \sim 1/\mu$ largely excludes anisotropic Rashba coupling as a source for that.

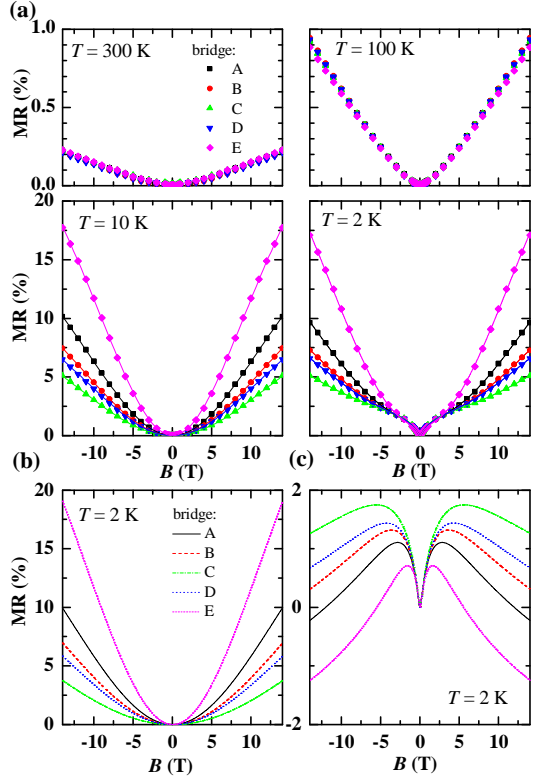

FIG. 5. (a) Magnetoresistance MR = $[R_s(B) - R_s(0)]/R_s(0)$ versus $B$ for the different microbridges of sample I recorded at $T$ = 300 K, 100 K, 10 K, and 2K. The solid lines are fits to the data with respect to the Kohler and Maekawa-Fukuyama expression (see text). (b) Contributions to MR at $T$ = 2 K from classical Lorentz scattering and (c) weak localization as deduced from the fitting parameters.

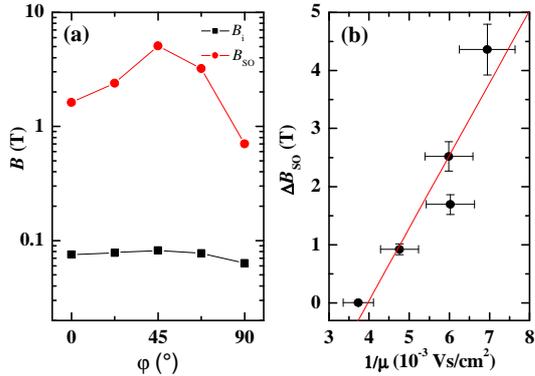

FIG. 6. (a) Inelastic field $B_i$ and spin orbit field $B_{SO}$ as a function of the in-plane direction of the current $0 \leq \varphi \leq 90°$, i. e., for bridge A – E of sample I. (b) The deviation $\Delta B_{so} = B_{so}(\varphi) - B_{so}(\varphi=90°)$ as a function of the inverse of the Hall mobility µ at 2 K. Linear fit to the data (solid line) and error bars are indicated.

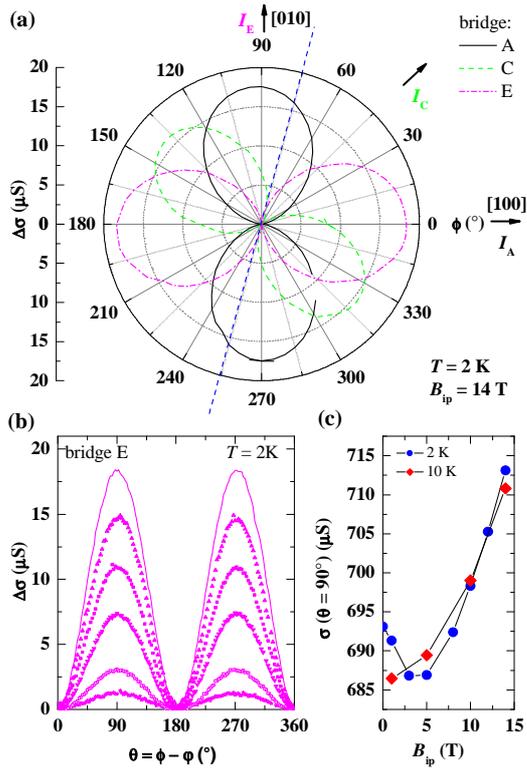

FIG. 7. (a) Polar plot of the magnetoconductance $\Delta\sigma(\phi) = \sigma(B,\phi) - \sigma(B,0)$ at $T = 2$ K and $B_{ip} = 14$ T for bridge A, C, and E of sample I as a function of $\phi$, the angle between the in-plane magnetic field $B_{ip}$ and the [100] direction. Crystallographic and current directions are indicated. Orientation of the step edges is shown by dashed blue line. (b) $\Delta\sigma$ of bridge E at $T = 2$ K versus $\theta$, the angle between the in-plane magnetic field $B_{ip}$ and the direction of current flow for various magnetic fields $B_{ip}$ (3, 5, 8, 10, 12, and 14 T – from bottom to top). The amplitude of $\Delta\sigma$ steadily increases with increasing $B_{ip}$. (c) Conductance $\sigma$ of bridge E with $B_{ip}$ perpendicular to current flow ($\theta = 90°$) versus $B_{ip}$ at $T = 2$ K and 10 K.

Applying the magnetic field parallel to the interface ($B_{ip}$) at an angle $\phi$ with respect to the [100] direction results in a strong field-induced anisotropy of the conductance, i. e., $\sigma = \sigma(\phi)$. Fig. 7 (a) displays a polar plot of the difference in conductance $\Delta\sigma = \sigma(B,\phi) - \sigma(B,0)$ for bridge A, C, and E of sample I as a function of the in-plane field direction. The measurements were carried out at $T = 2$ K and $B_{ip} = 14$ T. $\Delta\sigma(\phi)$ shows clear twofold anisotropic behavior with minima at $\phi \approx 0°/180°$, $45°/225°$, and $90°/270°$ for bridge A, C, and E, respectively. The anisotropy seems to depend much stronger on $\theta = \phi - \varphi$, the angel between $B_{ip}$ and the direction of current flow, than on the crystallographic direction or the orientation of the step edges. Small differences in amplitude are very likely related to differences in $R_s$ and $\mu$ of the microbridges. Plotting $\Delta\sigma$ versus $\theta$ (not shown here) indeed results in similar behavior of the microbridges, i. e., minima at $\theta = 0°/180°$ and maxima at $\theta = 90°/270°$. In Fig. 7 (b), $\Delta\sigma$ of bridge E is plotted as a function of $\theta$ at 2 K for various $B_{ip}$. With increasing field, the amplitude of $\Delta\sigma$ increases steadily reaching a value of $\approx$ 17 µS at $B_{ip}$ = 14 T corresponding to a relative change $\Delta\sigma/\sigma(B,0) \approx 2.6\%$. With increasing temperature, $\Delta\sigma$ decreases (not shown) falling below our measuring limit for $T > 50$ K.

The anisotropy in the longitudinal conductance $\sigma(B,\theta)$ likewise results in a modulation of the transverse conductance $\sigma_{xy}(B,\theta)$ with the same amplitude but a phase shifted by $\Delta\theta = 45°$ (not shown). The observed symmetric $\Delta\sigma_{xy}$ is a direct signature of the 2D anisotropic system [45].

Similar anisotropic behavior of $\Delta\sigma(\theta)$ was also found by other groups [40,46,45]. For in-plane magnetic field direction, $\sigma(\theta)$ is not affected by orbital contributions. Band-structure calculations on LAO/STO by Fête et al. [40] have shown, that the presence of a Rashba term in combination with the 1D-like $d_{xz}$ and $d_{yz}$ subbands - caused by the large difference of the electron mass along the two orthogonal in-plane directions - results in a spin-splitting and hence energy gap at the $\Gamma$ point if $B_{ip}$ is applied parallel to the current direction ($\theta = 0$), where the current direction is along the $x$- or $y$-direction. In contrast, $B_{ip}$ perpendicular to the current ($\theta = 90°$) only causes a Zeeman-like offset of the subbands. Therefore, Rashba induced magnetoconductance oscillations, i. e., $\Delta\sigma(\theta) \sim \sin^2(\theta)$ are only expected if charge transport is controlled by the $d_{xz}$, $d_{yz}$ subbands. In LAO/STO the band filling strongly depends on the sheet carrier density $n_s$ leading to a Lifshitz transition at $n_c \approx 1.7 \times 10^{13}$ cm$^{-2}$ [6] above which $d_{xz}$, $d_{yz}$ subbands become occupied. Note, $n_s \approx 2 - 3 \times 10^{13}$ cm$^{-2} > n_c$ for our AO/STO samples (*cf*. Fig. 4). In addition, since the $x$- and $y$-directions are orthogonal to each other, the degeneracy of the $d_{xz}$, $d_{yz}$ subbands results in the same anisotropy of $\Delta\sigma(\theta)$ for arbitrary current direction, as observed in our experiment. Therefore, the specific angular dependence of $\Delta\sigma(\theta)$ of AO/STO strongly suggest that anisotropic behavior is caused by Rashba spin-orbit interaction, too.

Under the simplified assumption that $\Delta\sigma(\theta)$ is entirely due to the $d_{xz}$, $d_{yz}$ subbands closest to the Fermi energy $\varepsilon_F$ the relative change $\Delta\sigma(B, \theta = 90°)/\sigma(B, 0) = 1/8(\Delta_{SO}/\varepsilon_F)^2$, is directly related to the Rashba induced spin-splitting energy $\Delta_{SO}$ [47]. For LAO/STO $\Delta\sigma(B,\theta = 90°)/\sigma(B,0) \approx 1.6\ \%$ at $B$ = 7 T, resulting in $\Delta_{SO}$ = 7 (2.5) meV for a zero-field conductance $\sigma_0$ = 2 (1) mS, respectively [3,40]. For AO/STO (sample I) $\Delta\sigma(B,$ $\theta = 90°)/\sigma(B, 0)$ increases from 0.2 % at $B = 3$ T via $\approx 1\%$ at 8 T to 2.6% at $B = 14$ T. Hence, $\Delta_{SO}$ seems to be smaller compared to LAO/STO.

Rashba-type SOC not only depends on $n_s$ [3] but also on the electric field at the interface [48] and hence polarity of the heterostructure. For the epitaxial grown spinel-type/perovskite heterostructure $\gamma$-$Al_2O_3$/STO the polar character and potential buildup is expected to be comparable to that of LAO/STO or even larger [49]. A low deposition temperature $T_s$, as used here, indeed leads to a strongly disordered and quasi amorphous structure of $Al_2O_3$. However, local residual polarity may still exist at the interface. Because of $n_s$ being equivalent to that of LAO/STO, we assume that AO/STO displays polar character alike, however, probably weaker compared to LAO/STO or $\gamma$-$Al_2O_3$/STO heterostructures.

The field-dependence of $\sigma$ for $B_{ip}$ perpendicular to the direction of current flow ($\theta = 90°$), is shown for $T = 2$ K and 10 K in Fig. 7 (c). At $T = 10$ K, $\sigma(B)$ steadily increases with increasing $B$ resulting in a negative MR of about 4% at 14 T. This is in stark contrast to the much higher and positive MR ($\approx 20\%$ at 14 T) for $B$ perpendicular to the interface. As discussed above, for in-plane magnetic field $\sigma(\theta)$ of the 2DES is not affected by orbital contributions. In addition, for $\theta = 90°$ a Zeeman-like offset of spin subbands emerges leading to spin-polarized bands. Hence with increasing $B$ interband scattering is suppressed, leading to a negative MR [50]. Interestingly, at $T = 2$ K $\sigma(B)$ first decreases, displaying a positive MR for $B_{ip} < 5$ T. The same behavior is also observed at low $T$ for LAO/STO when $n_s$ is close to $n_c$ which has been related to specific properties of the electronic band structure [40].

## IV. SUMMARY

Electronic transport of the 2D electron system in AO/STO heterostructures was investigated with respect to anisotropic behavior. To this purpose, microbridges with various in-plane orientations were patterned on (001) oriented $TiO_2$-terminated STO substrates displaying different step edge alignment. Below about 30 K $R_s$ displays significant anisotropy with respect to the in-plane direction of current flow. At low $T$ and $B = 0$ anisotropy, amounting up to 55%, is caused mainly by defect scattering and is hence non-crystalline in nature. Dominant contributions are suggested to result from anisotropic distribution of <110> dislocations in STO, being about 2 - 4 times larger in amplitude compared to anisotropic interfacial scattering by step edges. Anisotropic defect scattering likewise results in an anisotropic Hall mobility μ of the 2DES affecting magnetotransport. The main part of the normal MR originates from LZ scattering which becomes anisotropic via $w \sim 1/\mu$. For $T = 2$K, contributions from WL are apparent and dominate the total MR for $B < 2$ T. Interestingly WL shows anisotropic behavior as well. The spin-orbit field $B_{SO}$ displays specific behavior on φ and is the main reason for anisotropic MR for $T \leq 2$ K. The distinct relation $1/B_{SO} \sim \mu$ strongly motivates EY scattering mechanism underlining the important role of impurities and lattice defects and largely excludes anisotropic Rashba coupling as source of anisotropic WL. Applying the magnetic field parallel to the interface results in strong field-induced in-plane anisotropy of the conductance $\sigma(\theta) \sim \sin^2(\theta)$, with θ the angle between $B$ and the current direction. The anisotropy is very likely caused by Rashba spin-orbit interaction. Compared to LAO/STO, the Rashba SOC appears to be smaller in AO/STO which might be explained by a weaker polarity at the interface.

## ACKNOWLEDGEMENTS

Part of this work was supported by the Deutsche Forschungsgemeinschaft (DFG). We are grateful to R. Thelen and the Karlsruhe Nano Micro Facility (KNMF) for technical support.